\documentclass{ifacconf}

\usepackage{graphicx, amsmath, amssymb}      
\usepackage[table]{xcolor}
\usepackage[percent]{overpic}
\usepackage{natbib}        

\newcommand\ngrey{\cellcolor{black!10}}

\definecolor{myAs}{rgb}{0.85,0.325,0.098}
\definecolor{myA}{rgb}{0.301,0.745,0.933}

\newcommand{\pfcom}[1]{{#1}}

\begin{document}
\begin{frontmatter}

\title{Stability of cluster formations in adaptive Kuramoto networks}

\thanks[footnoteinfo]{This document contains a preprint version of the paper accepted for the presentation at the 24\textsuperscript{th} International Symposium on Mathematical Theory of Networks and Systems (MTNS 2021), Cambridge, United Kingdom.}
\thanks[footnoteinfo]{Funded by the Deutsche Forschungsgemeinschaft (DFG, German Research Foundation) -- Project-ID 434434223 -- SFB 1461.}

\author[First]{Petro Feketa} 
\author[First]{Alexander Schaum} 
\author[First]{Thomas Meurer}

\address[First]{Chair of Automatic Control, Kiel University, 24143 Kiel, Germany (e-mail: \{pf,alsc,tm\}@tf.uni-kiel.de)}

\begin{abstract}                
This paper studies stability properties of multi-cluster formations in Kuramoto networks with adaptive coupling. Sufficient conditions for the local asymptotic stability of the corresponding synchronization invariant toroidal manifold are derived and formulated in terms of the intra-cluster interconnection topology and plasticity parameters of the adaptive couplings. 
The proposed sufficient stability conditions qualitatively mimic certain  counterpart results for Kuramoto networks with static coupling which require sufficiently strong and dense intra-cluster connections and sufficiently weak and sparse inter-cluster ones. 
Remarkably, the existence of cluster formations depends on the interconnection structure between nodes belonging to different clusters and does not require any coupling links between nodes that form a cluster. On the other hand, the stability properties of clusters depend on the interconnection structure inside the clusters. This dependence constitutes the main contribution of the paper. Also, two numerical examples are provided to validate the proposed theoretical findings.
\end{abstract}


\end{frontmatter}

\section{Introduction}

Synchronization analysis of oscillator networks is an active research topic having a variety of applications in neurophysiology \citep{cattai2019phase, rohr2019frequency, menara2019framework}, distributed power generation \citep{balaguer2010control, berner2020adaptive} and power systems \citep{paganini2019global}, secure communication and chaos \citep{argyris2005chaos, FEKETA2019628}, memristive circuits~\citep{ignatov2016synchronization, ignatov2017memristive}, biochemical networks \citep{scardovi2010synchronization}, etc. A simple yet dynamically rich Kuramoto model proved to be an appropriate paradigm for synchronization phenomena \citep{acebron2005kuramoto, dorfler2014synchronization}.
Important results on frequency and phase synchronization can be found in monographs of \cite{boccaletti2008synchronized}, \cite{pikovsky2003synchronization}, \cite{strogatz2004sync}. Control theoretic approaches to the study of synchronization phenomena can be found in the works of \cite{DBSIAM}, \cite{jadbabaie2004stability},  \cite{chopra2009exponential}, \cite{lin2007state}, \cite{ scardovi2007synchronization}, \cite{schmidt2012frequency}, \cite{SCARDOVI20092557}. 

This paper aims at studying oscillator networks with adaptive coupling, i.e., networks whose coupling strengths between oscillators are driven by distinct dynamical systems in $\mathbb R$. This is motivated by the synaptic plasticity phenomenon observed in neurophysiological networks where the rate of signal propagation between neurons depends on their states \citep{abbott2000synaptic}. Together with the phases of oscillators which are defined on a circle, this leads to a dynamical system defined in the product of a multi-dimensional torus and Euclidean space. The emergence of synchrony in such networks corresponds to the existence of invariant manifolds of a lower dimension and of a special topological structure \citep{feketa2019synchronization}.

Recent results on full synchronization of adaptive and non-adaptive networks can be found in \cite{2018arXiv180807263Z}, \cite{jafarpour2018synchronization}, \cite{ha2016synchronization, ha2018emergent}, \cite{gushchin2016phase}. In particular, \cite{ha2016synchronization, ha2018emergent} provide sufficient conditions in terms of initial states for the full phase and frequency synchronization of adaptive networks with particular learning rules. Frequency synchronization of adaptive Kuramoto networks for some particular number of clusters is studied in \cite{gushchin2016phase}.

Phase multi-clustering is characterized by a partition of the network nodes into subsets where the nodes' phases evolve identically within each subset. Such subsets are called \emph{clusters}. For the case of static coupling, the emergence and stability of multi-cluster behaviors are studied in \cite{menara2019stability, menara2019exact, menara2019framework}, \cite{8263710}, \cite{4961065}, \cite{4961435}, \cite{XIA20112395}. Additionally, sufficient conditions in the form of algebraic connectivity for the partial synchronization of Kuramoto networks with static coupling have been recently presented in \cite{qin2019partial}. In \cite{5718119}, a class of phase models with state-dependent coupling that are gradients of suitable potential functions is identified. In \cite{feketa2019synchronization}, sufficient conditions are derived for the existence of multi-dimensional invariant toroidal manifolds which correspond to the multi-cluster behavior in Kuramoto networks with adaptive coupling.
In the present paper, new sufficient conditions for the local asymptotic stability of multi-cluster formations of adaptive Kuramoto networks are proposed, which are formulated in terms of the interconnection topology of the networks and plasticity parameters of adaptive coupling. By this, the paper  extends some results of \cite{menara2019stability} to the case of adaptive networks and complements the results of \cite{feketa2019synchronization} with sufficient conditions on the stability of multi-cluster formations. These new sufficient conditions provide qualitative relations between the intra- and inter-cluster network connectivity and the plasticity parameters of the corresponding adaptive couplings.

Contrary to the notion of phase multi-clustering used in this paper, in \cite{berner2018multi, berner2019self}, the authors derive conditions for the frequency clustering in adaptive networks of identical Kuramoto oscillators and study the possible arrangement of phases within every cluster. Existence criteria for multi-cluster solutions, where different clusters correspond to different frequencies, and their explicit form are presented in these works.

The paper is organized as follows. In Section~2, the main object of investigation is presented and some known results on the existence of invariant toroidal manifolds, which correspond to the multi-cluster behavior of the adaptive Kuramoto networks, are recalled. In Section~3, new sufficient stability conditions for the desired multi-cluster formations are derived and their interpretation is given. The proof of the main result is based on the perturbation theory of invariant tori of dynamical systems defined in the product of a torus and a Euclidean space \citep{SAMOILENKO19973121}. Numerical examples in Section~4 and a short discussion in Section~5 conclude the paper.

\subsubsection{Notation.}
Let $\mathbb N${\color{black}, $\mathbb R$,} and $\mathbb R_{>0}$ denote the sets of natural{\color{black}, real,} and positive real numbers, respectively. For given $n,m\in\mathbb N$ let $\mathbb R^n$ and $\mathcal T_m$ denote the $n$-dimensional Euclidean space and $m$-dimensional torus, respectively. {\color{black}The one-dimensional torus $\mathcal T_1$ is the one-sphere (circle).} Let $f:\mathcal T_m \to \mathbb R^n$ be a function of the variable $\varphi=(\varphi_1,\ldots,\varphi_m)^\top\in\mathcal T_m$ which is continuous and $2\pi$-periodic with respect to each $\varphi_s$, $s=\overline{1,m}$. Finally, $C(\mathcal T_m)$ denotes the space of all such functions $f$ equipped with the norm
$
|f|_0 = \max_{\varphi \in \mathcal T_m} \left\|f(\varphi)\right\|,
$
where $\left\|\cdot\right\|$ denotes the Euclidean norm in $\mathbb R^n$, i.e., $\left\|f(\varphi)\right\|^2=\sum_{i=1}^n |f_i(\varphi)|^2$, $|f_i(\varphi)|$ stands for the absolute value of the $i$-th component of $f$ evaluated at $\varphi$.
By $C^1(\mathcal T^m)$ we denote the subspace of $C(\mathcal T_m)$ with every $f\in C^1(\mathcal T_m)$ having a continuous partial derivative with respect to each $\varphi_s$, $s=\overline{1,m}$ and
$
|f|_1 = \max \left\{|f|_0, \left|\frac{\partial f}{\partial \varphi_1}\right|_0, \ldots, \left|\frac{\partial f}{\partial \varphi_m}\right|_0 \right\}.
$
$\operatorname{Re}\lambda(A)$ denotes the set of real parts of all eigenvalues of square matrix $A$ and any set $B<0$ if and only if for any $b\in B$ it holds that $b<0$.

\section{Preliminaries and Problem Statement}

Let ${\color{black}\mathcal G} = (\mathcal V, \mathcal E)$ be the directed graph representing the network of oscillators, where $\mathcal V=\{1,\ldots,N\}$ and $\mathcal E \subseteq \mathcal V \times \mathcal V$ represent the oscillators and their interconnection edges, respectively. Let $A = [a_{ij}]_{i,j=\overline{1,N}}$ be the adjacency matrix of ${\color{black}\mathcal G}$, where $a_{ij}=1$ if the edge $(i,j)\in\mathcal E$, and $a_{ij}=0$ when $(i,j)\not\in\mathcal E$. Additionally, it is assumed that the graph does not have self-loops, i.e., $a_{ii}=0$ for all $i=\overline{1,N}$. The dynamics of the network is given by
\begin{equation}\label{main1}
\begin{array}{rcll}
\dot \theta_i & = & w_i+\sum\limits_{j=1}^N a_{ij} k_{ij} \sin(\theta_j-\theta_i), & i=\overline{1,N}, \\
\dot k_{ij} &=& -\gamma k_{ij} + \mu_{ij}\Gamma(\theta_j-\theta_i), & i,j=\overline{1,N},
\end{array}
\end{equation}
where $w_i\in\mathbb R$ and $\theta_i(t)\in {\color{black}\mathcal T_1}$ denote the natural frequency and the phase of the $i$-th oscillator. The dynamics of the coupling strength $k_{ij}(t)\in\mathbb R$ is defined by parameters $\mu_{ij}, \gamma \in \mathbb R_{>0}$ and $\Gamma\in C^1(\mathcal T_{\pfcom{1}})$ with $|\Gamma|_1=\delta\in\mathbb R_{>0}$.

The network exhibits cluster synchronization when the oscillators can be partitioned into subsets so that the phases of the oscillators in each subset evolve identically. This type of behavior corresponds to the existence of an invariant toroidal manifold of \pfcom{the corresponding error-}system \pfcom{\eqref{main2-torus}-\eqref{main2-intra}}, see \cite{feketa2019synchronization} \pfcom{for details}.

Let $\mathcal P = \{\mathcal P_1,\ldots, \mathcal P_m\}$ with $m\in\mathbb N$, $m>1$, be a partition of $\mathcal V$, where $\cup_{i=1}^{m}\mathcal P_i =\mathcal V$ and $\mathcal P_i \cap \mathcal P_j = \varnothing$ if $i\not=j$. For a given partition $\mathcal P$, let $\mathcal E_{in}$ and $\mathcal E_{out}$ be the subsets of $\mathcal E$ that correspond to the intra-cluster links and inter-cluster links, respectively. The cardinalities of these sets
\begin{equation}\label{Eq:cardinality}
c_{in}=\mathtt{card}\{\mathcal E_{in}\} \quad \text{and} \quad c_{out}=\mathtt{card}\{\mathcal E_{out}\}
\end{equation}
characterize the interconnection structure of $\mathcal G$ with respect to the partition $\mathcal P$. Additionally, let
\begin{equation*}
w_{min}=\min\limits_{i=\overline{1,N}} |w_i| \quad \text{and}\quad w_{max}=\max\limits_{i=\overline{1,N}} |w_i|
\end{equation*}
denote the minimal and maximal \pfcom{absolute value of} natural frequencies.

Sufficient conditions for the existence and construction procedure of an  invariant toroidal manifold that corresponds to the $m$-cluster behavior of the network \eqref{main1} have been proposed in \cite{feketa2019synchronization} for the case of identical plasticity parameters $\mu_{ij}\equiv \mu\in\mathbb R_{>0}$. However, the mentioned result does not answer the question whether the constructed invariant manifold is (asymptotically) stable. The present paper answers this question and proposes sufficient conditions for the asymptotic stability of the invariant manifold for the case when the plasticity parameters $\mu_{ij}$ are different for the intra-cluster and inter-cluster links. Namely, let $\mu_{ij}\equiv\tilde\mu\in\mathbb R_{>0}$ if the link $(i,j)$ connects nodes within some cluster $\mathcal P_s$, $s=\overline{1,m}$, and $\mu_{ij}\equiv\mu\in\mathbb R_{>0}$ otherwise. For convenience purposes, system \eqref{main1} with the chosen set of plasticity parameters will be denoted as $\Sigma(\tilde \mu, \mu)$ from now on.

\pfcom{Let partition $\mathcal P$ be given. For every cluster $\mathcal P_s$, $s=\overline{1,m}$ pick an arbitrary oscillator $i_s \in \mathcal P_s$ and denote its phase and natural frequency by $\varphi_s:=\theta_{i_s}$ and $\bar w_s:=w_{i_s}$, respectively.
For every oscillator $i\in\mathcal P_s$, let $e_i=\theta_i-\varphi_s$ define the relative phase-error within a given cluster $\mathcal P_s$, $s=\overline{1,m}$. Then, $\Sigma(\tilde \mu, \mu)$ can be rewritten in the following form:
\begin{align}
\notag
\dot\varphi_s  =  &\bar w_s +
\sum\limits_{j\in\mathcal P_s}a_{i_sj} k_{i_sj}\sin{e_j} \\ 
\label{main2-torus}
&+\sum\limits_{r\not=s}\sum\limits_{j\in\mathcal P_r} a_{i_sj}k_{i_sj}\sin(e_j+\varphi_r-\varphi_s),\, s=\overline{1,m},
\end{align}
\begin{align}
\notag
\dot e_i  =  &w_i-\bar w_s + \sum\limits_{j\in\mathcal P_s}\left[a_{ij} k_{ij}\sin(e_j-e_i)-a_{i_sj} k_{i_sj}\sin{e_j}\right] &~  \\
\notag
&+\sum\limits_{r\not=s}\sum\limits_{j\in\mathcal P_r} \left[a_{ij} k_{ij}\sin(e_j-e_i+\varphi_r-\varphi_s)\right. &~\\
\notag
&\qquad\qquad\quad\left.- a_{i_sj} k_{i_sj}\sin(e_j+\varphi_r-\varphi_s) \right]\\
\label{main2-errors}
&\qquad\qquad\qquad\qquad\qquad\forall i\in \mathcal P_s\setminus\{i_s\},\, s=\overline{1,m},\\[3mm]
\notag
\dot k_{ij}  = &-\gamma k_{ij} + \mu \Gamma(e_j-e_i+\varphi_r-\varphi_s) \\
\label{main2-inter}
&\qquad\qquad\quad\forall i\in\mathcal P_s,\,\, \forall j\in\mathcal P_r,\, s\not=r, s,r=\overline{1,m},\\[3mm]
\label{main2-intra}
\dot k_{ij}  = &-\gamma k_{ij} + \tilde\mu \Gamma(e_j-e_i)\, \,\,\, \forall i,j\in\mathcal P_s,\, i\mathord{\not=}j,\, s\mathord{=}\overline{1,m}.
\end{align}
System \eqref{main2-torus}-\eqref{main2-intra} has the same number of equations as system \eqref{main1}. Equations \eqref{main2-torus} describe the dynamics of $m$ arbitrarily selected oscillators (one from every cluster). Equations \eqref{main2-errors} describe the error dynamics within each cluster. Equations \eqref{main2-inter} describe the dynamics of the coupling strengths between nodes of different clusters. Finally, \eqref{main2-intra} describe the dynamics of the intra-cluster coupling strengths. Let $\varphi = (\varphi_1, \ldots, \varphi_m)^{\top}\in\mathcal T_m$ and $e=(e_{i^1_1},\ldots,e_{i^m_{n_m}})\in\mathbb R^{N-m}$ be the vectors collecting all cluster phases $\varphi_i$, $i=\overline{1,m}$ and all intra-cluster relative phase errors $e_i$, $i\in \mathcal P_s\setminus\{i_s\}$, $s=\overline{1,m}$, respectively.
Similarly, all inter- and intra-cluster coupling strengths are collected into the vectors $k^{inter}\in\mathbb R^{c_{out}}$ and $k^{intra}\in\mathbb R^{c_{in}}$, respectively, and $k=({k^{inter}}^\top, {k^{intra}}^\top)^\top$.
The multi-cluster behavior in network $\Sigma(\tilde \mu, \mu)$ is possible if system \eqref{main2-torus}-\eqref{main2-intra} possesses an invariant toroidal manifold
\begin{equation}\label{main3}
e \equiv  0,\quad  k = u(\varphi), \quad \varphi\in\mathcal T_m
\end{equation}
for some $u\in C(\mathcal T_m)$. This invariant manifold corresponds to the oscillating behavior of coupling strengths $k$ preserving zero phase error $e$ within clusters. In \cite{feketa2019synchronization}, it has been shown that the inter-cluster coupling strengths cannot converge to some constant value say $d$ simultaneously guaranteeing the convergence of the phase errors to zero, i.e., $e=0$, $k=d$, $\varphi\in\mathcal T_m$ is \emph{not} an invariant set of \eqref{main2-torus}-\eqref{main2-intra} for any constant $d\in\mathbb R^{c_{in}+c_{out}}$. Hence, the oscillating behavior of the inter-cluster coupling strengths is necessary for the emergence of multi-cluster formations in \eqref{main1}.
}
\pfcom{F}ollowing the steps of the proof of Theorem~3 from \cite{feketa2019synchronization}, the following proposition can be concluded.
\begin{thm}[adapted from \cite{feketa2019synchronization}, Thm.~3]\label{thm_main}
Let the following conditions hold true for system $\Sigma(\tilde \mu,\mu)$ and a given partition $\mathcal P$:
\begin{itemize}
\item[(A1)] for any $s=\overline{1,m}$ and for any $i,j\in \mathcal P_s$ $$w_i=w_j;$$
\item[(A2)] for any $s,r=\overline{1,m}$, $s\not =r$ there exist constants $c_{sr}\in\mathbb N$ such that for any $i\in\mathcal P_s$ $$\sum\limits_{j\in\mathcal P_r} a_{ij}=c_{sr};$$
\item[(A3)] given $c_{max}:=\max\limits_{s=\overline{1,m}} \sum\limits_{r\not=s}c_{sr}$ it holds that
\begin{equation}\label{eqA31}
w_{min}-\mu\gamma^{-1}\delta c_{max}>0
\end{equation}
and
\begin{equation}\label{eqA32}
4\frac{\mu}{\gamma^2} \delta \sqrt{c_{out}} \sum\limits_{\substack{s,r=\overline{1,m} \\ s\not = r}}{c_{sr}} \frac{w_{max}+\mu\gamma^{-1}\delta c_{max}}{w_{min}-\mu\gamma^{-1}\delta c_{max}}<1.
\end{equation}
\end{itemize}
Then, system \pfcom{\eqref{main2-torus}-\eqref{main2-intra}} has an invariant toroidal manifold, which corresponds to the $m$-cluster behavior \pfcom{of $\Sigma(\tilde \mu, \mu)$} defined by the partition $\mathcal P$.
\end{thm}

\subsubsection{Remark 1.}
Conditions (A1)-(A3) allow for the following interpretation:
\begin{itemize}
\item (A1) requires the natural frequencies to be equal within every cluster.
\item (A2) requires that the number of links coming to every node
  within a given cluster $\mathcal P_s$ from other given cluster
  $\mathcal P_r$, $r\not=s$ is the same. The number of incoming links to the nodes of $\mathcal P_s$ from the cluster other than $\mathcal P_r$ may be different. Also, (A2) restricts only the number of links and does not require any symmetry of the corresponding adjacency matrix. It is worth to highlight that the intra-cluster couplings are generally not required for the emergence of multi-cluster behavior in the network. This type of behavior may result from a proper interaction of nodes with the nodes of other clusters. In Theorem~\ref{thm-stab}, which is the main result of this paper, it will be shown that the intra-cluster links are an essential ingredient for the stability of clusters.
\item (A3) establishes the relations between the natural frequencies of the oscillators, plasticity parameters $\mu, \gamma, \delta$ and the inter-cluster interconnection topology. For a given network of Kuramoto oscillators and a given partition $\mathcal P$ conditions (A3) can always be satisfied by choosing a sufficiently small plasticity parameter $\mu$.
\end{itemize}

\section{Stability of multi-cluster formations}

In this section, sufficient conditions for the asymptotic stability of multi-cluster formations are derived.
For a given cluster $\mathcal P_s$, $s=\overline{1,m}$, let
\begin{itemize}
\item $n_s$ be the number of elements in the set $\mathcal P_s$;
\item $\mathcal G_s\subset \mathcal G$ be a subgraph that correspond to the nodes from $\mathcal P_s$ and intra-cluster connections, i.e.,
$$\mathcal G_s = \{(\mathcal P_s, \mathcal E_s): \mathcal E_s = \mathcal P_s \times\mathcal P_s \cap \mathcal E\};$$
\item $A_s$ be the adjacency matrix of $\mathcal G_s$.
\end{itemize}

In order to define the residual connectivity of $\mathcal G_s$ with respect to the node $i_s$, the nodes inside each cluster are enumerated according to the rule $\mathcal P_s = \{i^s_1, \ldots, i^s_{k_s}, \ldots, i^s_{n_s}\}$, where $i^s_{k_s}=i_s$, i.e., the selected node $i_s$ has a sequential number $k_s$ in the cluster $\mathcal P_s$. Then, let
\begin{itemize}
\item $A_s^{-}$ be an $(n_s-1)\times (n_s-1)$-dimensional matrix constructed from $A_s$ by removing its $k_s$-th row and column; 
\item $\tilde A_s$ be the residual adjacency matrix w.r.t. the node $i_s$, i.e.,
\end{itemize}
\begin{equation}\label{auxold}
\begin{split}
\tilde A_s = A_s^- -
\begin{pmatrix}
a_{i_s i^s_1} & \ldots & a_{i_s i^s_{k_s-1}} & a_{i_s i^s_{k_s+1}} & \ldots & a_{i_s i^s_{n_s}} \\
a_{i_s i^s_1} & \ldots & a_{i_s i^s_{k_s-1}} & a_{i_s i^s_{k_s+1}} & \ldots & a_{i_s i^s_{n_s}} \\
\vdots & ~ &\vdots & \vdots& ~ & \vdots \\
a_{i_s i^s_1} & \ldots & a_{i_s i^s_{k_s-1}} & a_{i_s i^s_{k_s+1}} & \ldots & a_{i_s i^s_{n_s}}
\end{pmatrix}
\end{split}
\end{equation}
\begin{itemize}
\item $D_s$ be the degree matrix of $A_s$, i.e., the diagonal matrix with diagonal elements equal to the sum of all elements in the corresponding row of $A_s$, and $D_s^-$ be an $(n_s-1)\times (n_s-1)$-dimensional matrix constructed from $D_s$ by removing its $k_s$-th row and column.
\end{itemize}

\begin{thm}\label{thm-stab}
Let network $\Sigma(\tilde \mu, \mu)$ satisfy conditions (A1), (A2), and (A3) of Theorem~1, i.e., there exist the synchronization invariant toroidal manifold, which corresponds to the multi-cluster behavior of the network given by partition $\mathcal P$. If
\begin{itemize}
\item[(A4)] for every $s=\overline{1,m}$
\begin{equation}
\operatorname{sign}\Gamma(0) \operatorname{Re} \lambda\left(\tilde A_s - D_s^- \right)<0,
\end{equation}
\end{itemize}
then there exist $\mu_0\leq\mu$ such that for all $\nu< \mu_0$ the invariant toroidal manifold \pfcom{that corresponds to the multi-clustering of} $\Sigma(\tilde \mu, \nu)$ is locally asymptotically stable. 
\end{thm}
\begin{pf}
Due to space limitations, only the main steps of the proof are presented. The reader may consult the monographs \cite{Sam1, mitr} and the papers \cite{SAMOILENKO19973121, PF14} for the concepts of \emph{nonlinear extension of dynamical system on torus} and \emph{Green-Samoilenko function of the invariant tori problem}. The relation of these notions to the synchronization analysis of oscillator networks is also discussed in \cite{feketa2019synchronization}.

\pfcom{A}ssume that the conditions of Theorem 1 hold true and the invariant toroidal manifold \eqref{main3} exists. Moreover, from~\cite{feketa2019synchronization}, it follows that this invariant torus has the form
\begin{align}\notag
e\equiv 0, \quad k^{intra}&=\left(\frac{\tilde\mu\Gamma(0)}{\gamma}, \ldots, \frac{\tilde\mu\Gamma(0)}{\gamma}\right)^\top=:u^{intra}(\varphi),\\ \label{endtor}  k^{inter}&=u^{inter}(\varphi), \quad \varphi\in\mathcal T_m,
\end{align}
and allows for the estimates
\begin{equation}\label{torest}
|u^{intra}|_0\leq \frac{\tilde\mu}{\gamma}\delta\sqrt{c_{in}} \quad \text{and} \quad |u^{inter}|_0\leq \frac{\mu}{\gamma}\delta\sqrt{c_{out}}  .
\end{equation}
In particular, this means that by a suitable choice of plasticity parameter $\mu$ one can make the oscillation amplitudes of the inter-cluster coupling strengths arbitrarily small.

For the purpose of (local) stability analysis of the invariant torus \eqref{endtor}, system \eqref{main2-torus}-\eqref{main2-intra} is considered in the domain $\mathcal D = \{(e,k,\varphi)\in \mathbb R^{N-m+c_{out}+c_{in}}\times \mathcal T_m: |e|\leq h\}$ for some (possibly small) $h\in \mathbb R_{>0}$.

System \eqref{main2-torus}-\eqref{main2-intra} can be rewritten in the following form
\begin{align}
\label{stability-eq1}
\dot \varphi &= \bar w + a(k^{intra},e)+b(k^{inter}, e, \varphi),\\
\label{stability-eq2}
\dot e &= f(k^{intra},e)+g(k^{inter}, e, \varphi), \\
\label{stability-eq3}
\dot k^{inter} &= -\gamma I k^{inter}+\mu G_1(e,\varphi), \\
\label{stability-eq4}
\dot k^{intra} &= -\gamma I k^{intra}+\tilde\mu G_2(e),
\end{align}
with appropriately chosen functions $f$ and $a, b, g, G_1, G_2$, where the latter ones will be considered as perturbation terms from now on.

The perturbation terms $a, b, g$, and $\mu \mathcal G_1$ can be made as small as needed by picking sufficiently small $\mu$ and sufficiently small constant $h$ \pfcom{that} defines the admissible domain for the errors $|e|<h$. In  particular, small constant $\mu$ means that all components $b_s$, $s=\overline{1,m}$ and $g_i$, $i\in\mathcal P_s\setminus\{i_s\}$, $s=\overline{1,m}$ of the corresponding perturbations $b$ and $g$
\begin{align}
\notag
b_s(k^{inter}, e, \varphi) = \sum\limits_{r\not=s}\sum\limits_{j\in\mathcal P_r} &a_{i_sj}k_{i_sj}\sin(e_j+\varphi_r-\varphi_s),
\\ \notag
g_i(k^{inter}, e, \varphi) = \sum\limits_{r\not=s}\sum\limits_{j\in\mathcal P_r} &\big[a_{ij} k_{ij}\sin(e_j-e_i+\varphi_r-\varphi_s)\\\label{auxnew}
- a_{i_sj} &k_{i_sj}\sin(e_j+\varphi_r-\varphi_s) \big]
\end{align}
are small since every $k_{* j}$ from \eqref{auxnew} is sufficiently small in a small vicinity of the invariant manifold \eqref{endtor} due to \eqref{torest}. Each component $a_s$, $s=\overline{1,m}$ of the perturbation $a$ 
\begin{equation*}
a_s(k^{intra},e)=\sum\limits_{j\in\mathcal P_s}a_{i_sj} k_{i_sj}\sin{e_j}
\end{equation*}
can be made arbitrary small by considering sufficiently small $h$, which defines a vicinity of the invariant manifold in $\mathcal D$, i.e., $|e|<h$.

Thanks to the conditions (A1)--(A3) and following \eqref{endtor}, equation \eqref{stability-eq4} has a fixed point $e=0$, $k^{intra}=\left(\frac{\tilde\mu\Gamma(0)}{\gamma}, \ldots, \frac{\tilde\mu\Gamma(0)}{\gamma}\right)^\top$ and the evolution of intra-cluster coupling strengths is independent of the evolution of the clusters' phases $\varphi$.

Then, the stability properties of the invariant toroidal manifold of \eqref{stability-eq1}-\eqref{stability-eq4} are determined by the properties of the Green-Samoilenko function of the corresponding unperturbed system \cite[Sect IV, \S 1-3]{Sam1}
\begin{align}
\label{stability-eq1u}
\dot \varphi &= \bar w,\\
\label{stability-eq2u}
\dot e &= f(k^{intra},e), \\
\label{stability-eq3u}
\dot k^{inter} &= -\gamma I k^{inter}. 
\end{align}
Since the right-hand sides of equations \eqref{stability-eq2u}, \eqref{stability-eq3u} do not depend on $\varphi\in \mathcal T_m$, the Green-Samoilenko function of \eqref{stability-eq1u}-\eqref{stability-eq3u} (which defines the invariant manifold for \eqref{main2-torus}-\eqref{main2-intra} and guarantees its asymptotic stability) exists if the corresponding linearization of $f$ in the vicinity of the invariant torus is exponentially stable \cite[Sect. III, \S 5]{Sam1}. In this case, sufficiently small perturbations of the right-hand side do not destroy the invariant manifold. As it has been discussed previously, the smallness of the perturbations can be guaranteed by choosing a sufficiently small plasticity parameter $\mu$ and looking into a sufficiently small vicinity of the invariant manifold.

Next, the linearization of $f$ in a small vicinity of the invariant manifold will be studied. Since for each $i\in\mathcal P_s\setminus\{i_s\}$, $s=\overline{1,m}$
$$f_i=\sum\limits_{j\in\mathcal P_s}\left[a_{ij} k_{ij}\sin(e_j-e_i)-a_{i_sj} k_{i_sj}\sin{e_j}\right]$$ depends only on the intra-cluster errors within cluster $\mathcal P_s$, the partial derivative $\frac{\partial f}{\partial e}$ is a block-diagonal matrix consisting of $m$ blocks ($m$ is the number of clusters). The entries of every block are given by
\begin{equation*}
\frac{\partial f_i(k^{intra},e)}{\partial e_i} = \sum_{j\in\mathcal P_s}\left[-a_{ij}k_{ij}\cos(e_j-e_i)\right]-a_{i_si}k_{i_si}\cos(e_i)
\end{equation*}
for $i\in\mathcal P_s \setminus \{i_s\}$ and
\begin{equation*}
\frac{\partial f_i(k^{intra}),e)}{\partial e_z} = a_{iz}k_{iz}\cos(e_z-e_i)-a_{i_sz}k_{i_sz}\cos(e_z)
\end{equation*}
for $z\in\mathcal P_s \setminus \{i_s\}$, $z\not =i$.

Evaluation of the partial derivative at $e=0$, $k^{intra}=\frac{\tilde\mu\Gamma(0)}{\gamma}$ leads to
\begin{equation}\label{auxaux}
\begin{split}
\frac{\partial f_i\left(\frac{\tilde\mu\Gamma(0)}{\gamma},0\right)}{\partial e_i} &= \frac{\tilde\mu\Gamma(0)}{\gamma} \Bigg(-\underbrace{\color{black}\sum_{j\in\mathcal P_s}a_{ij}}_{(*)}-\underbrace{\color{black}a_{i_si}}_{(**)}\Bigg), \\ \frac{\partial f_i\left(\frac{\tilde\mu\Gamma(0)}{\gamma},0\right)}{\partial e_z} &= \frac{\tilde\mu\Gamma(0)}{\gamma}\big(\underbrace{\color{black}a_{iz}}_{(***)}-\underbrace{\color{black}a_{i_sz}}_{(**)}\big)
\end{split}
\end{equation}
for $i,z\in\{i^s_1,\ldots,i^s_{k-1},i^s_{k+1},\ldots, i^s_{n_s}\}$, $i\not=z$. In \eqref{auxaux}, $(*)$ corresponds to the degree matrix $D^-_s$, $(\mathord{*}\mathord{*}\mathord{*})$ corresponds to the adjacency matrix $A_s^-$, and terms $(**)$ correspond to the auxiliary matrix used in \eqref{auxold}. Hence,
\begin{equation*}
\frac{\partial f\left(\frac{\tilde\mu\Gamma(0)}{\gamma},0\right)}{\partial e} = \frac{\tilde\mu\Gamma(0)}{\gamma} \text{diag}\left\{\tilde A_1-D^-_1, \ldots, \tilde A_m-D^-_m \right\}.
\end{equation*}

Condition (A4) implies that the real parts of all eigenvalues of $\frac{\partial f\left(\frac{\mu\Gamma(0)}{\gamma},0\right)}{\partial e}$ are negative. Summarizing, under conditions of Theorem~\ref{thm-stab}, system \eqref{stability-eq1}-\eqref{stability-eq4} can be represented in the form
\begin{equation*}
\begin{split}
\dot \varphi &= \bar w + \xi_1(\varphi, e, k^{inter}, k^{intra}), \\
\begin{pmatrix}
\dot e \\ \dot k^{inter} \\ \dot k^{intra}
\end{pmatrix}
&= (H+\xi_2(\varphi, e, k))\begin{pmatrix}
e \\ k^{inter} \\ k^{intra}
\end{pmatrix}
+ 
\begin{pmatrix}
0 \\ \mu G_1(0,\varphi) \\ \tilde \mu G_2(0)
\end{pmatrix},
\end{split}
\end{equation*}
where $H$ is a Hurwitz matrix and $\xi_1$, $\xi_2$ can be made arbitrarily small by appropriate choice of $h$ and $\mu$.
This is sufficient for the asymptotic stability of the  invariant toroidal manifold \eqref{endtor} of \eqref{stability-eq1}-\eqref{stability-eq4} for sufficiently small plasticity parameter $\mu$. This completes the proof.$\hfill\qed$
\end{pf}

For a suitable choice of plasticity parameters and natural frequencies of oscillators (see (A1) and (A3)), two additional structural conditions (A2) and (A4) are essential for the existence and stability of the desired multi-cluster formation. Condition (A2) restricts the interconnection structure between different clusters and is sufficient for the existence of the corresponding invariant toroidal manifold. Condition (A4) restricts the interconnection structure inside clusters and is sufficient for the asymptotic stability of clusters. However, (A4) is not necessary for the existence. The threshold $\mu_0$ for the plasticity parameter $\mu$ in Theorem~\ref{thm-stab} can be influenced by adjusting the intra-cluster interconnection topology. For example, $\mu_0$ can be enlarged by pushing the eigenvalues of the matrices $\tilde A_s - D_s^-$, $s=\overline{1,m}$ further to the left (from the definition of $\tilde A_s - D_s^-$, it can be seen that its eigenvalues depend on intra-cluster interconnection topology only). On the other hand, the same effect can be reached by increasing the plasticity parameter $\tilde\mu$, which directly influences the absolute value of the eigenvalues of $\frac{\partial f}{\partial e}$. Large $\tilde \mu$ leads to strong intra-cluster connections (see \eqref{endtor} and \eqref{torest}). This observation qualitatively mimics a recent result in \cite{menara2019stability} for the Kuramoto networks with static coupling, where strong intra-cluster couplings and weak inter-cluster couplings are required for the local asymptotic stability of the multi-cluster formation. Theorem~\ref{thm-stab} also suggests that the stability of multi-cluster formations can be concluded in the case of weak both intra- and inter-cluster couplings, assuming an appropriate interconnection structure inside the clusters so that all eigenvalues of $\tilde A_s - D_s^-$, $s=\overline{1,m}$ are in the left-half plane and sufficiently far away from zero.

\subsubsection{Remark~3 (fully connected network).} In the case of all-to-all connections, the structural condition (A2) for the existence of the invariant manifold is satisfied. By a direct calculation one may check that for the \pfcom{complete digraph $\mathcal G$} matrix $\tilde A_s - D_s^-$ for every $s=\overline{1,m}$ is a diagonal Hurwitz matrix with one real negative eigenvalue $-n_s$ of algebraic multiplicity $n_s-1$. Hence, the fulfillment of (A4) depends only on the sign of $\Gamma(0)$, which should be positive.

\subsubsection{Remark~4.} Condition (A4) is a clusterwise requirement meaning that it can be verified for each cluster separately. If condition (A4) fails for some cluster $\mathcal P_s$, this means that Theorem 2 cannot be used to conclude the asymptotic stability of this particular cluster, but all other clusters for which (A4) holds are asymptotically stable.

\section{Numerical examples}

The usage of Theorem 2 will be illustrated on two examples of adaptive Kuramoto networks from \cite{feketa2019synchronization} for which the conditions of Theorem 1 hold true, i.e., the desired $m$-cluster behavior is possible. Here it will be additionally checked whether this behavior is attractive.

\subsubsection{Example 1.}
\begin{figure}[!ht]
\begin{center}
\begin{overpic}[width=0.25\textwidth]{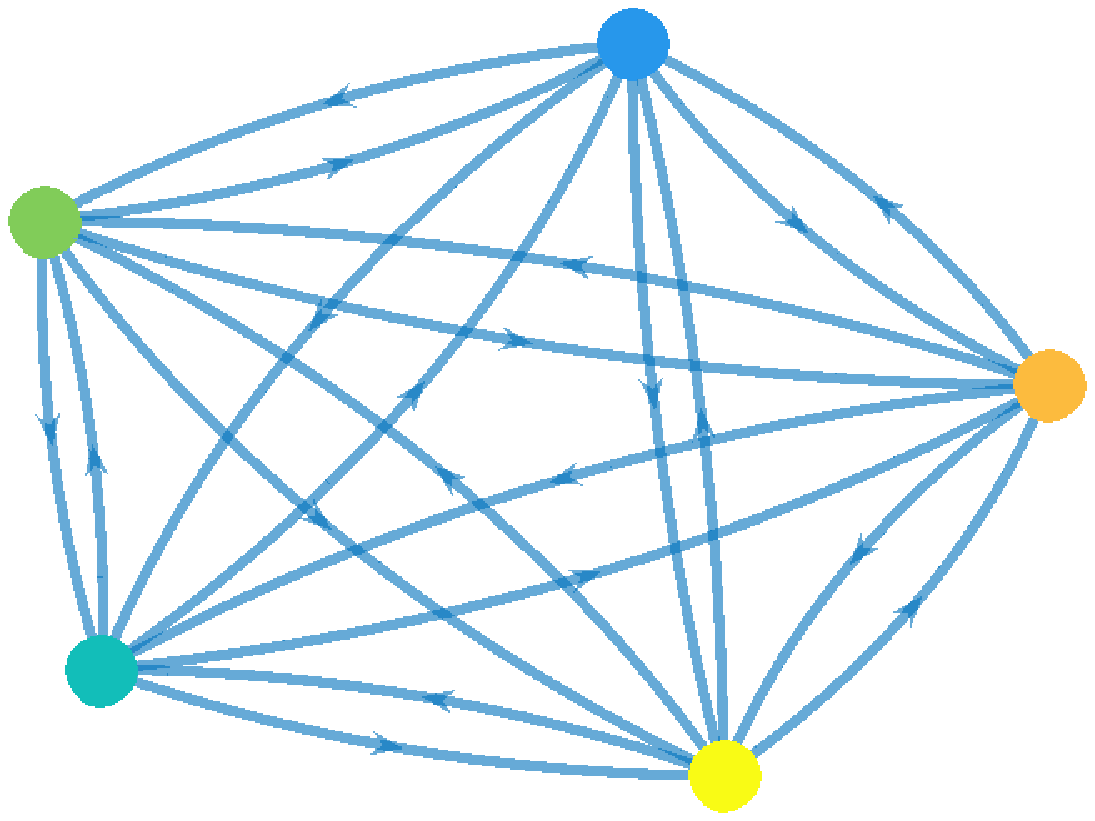}
 \put (57.8,60.3) {\tiny$\displaystyle 1$}
 \put (21.2,17.4) {\tiny$\displaystyle 2$}
 \put (17.5,48.2) {\tiny$\displaystyle 3$}
 \put (86.3,36.6) {\tiny$\displaystyle 4$}
 \put (64.0,10.0) {\tiny$\displaystyle 5$}
\end{overpic}\hspace{-4mm}
\begin{overpic}[width=0.25\textwidth]{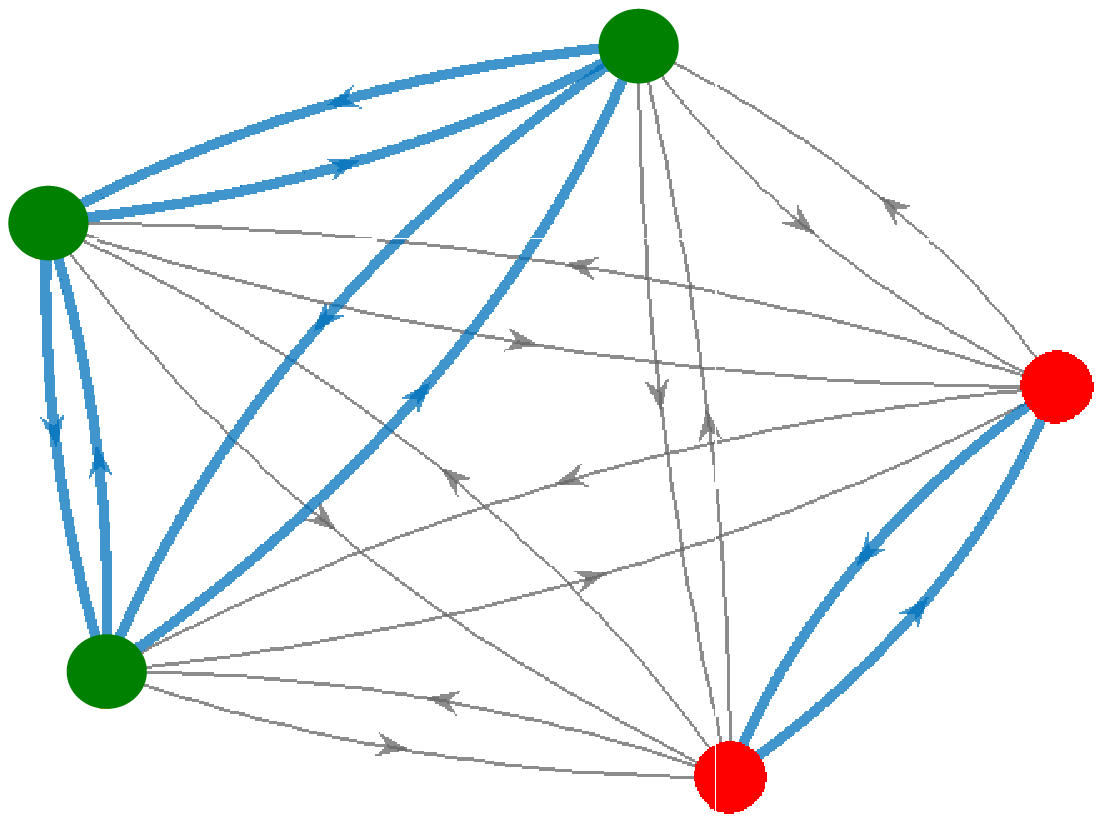}
 \put (57.6,60.3) {\tiny$\displaystyle 1$}
 \put (21.2,17.4) {\tiny$\displaystyle 2$}
 \put (17.0,48.2) {\tiny$\displaystyle 3$}
 \put (86.3,36.6) {\tiny$\displaystyle 4$}
 \put (64.0,10.0) {\tiny$\displaystyle 5$}
 \put (0,35) {\Large$\displaystyle \Rightarrow$}
\end{overpic}
\end{center}
\caption{(Example~1) {Visualization of the graph $\mathcal G$ and coupling strengths} at the beginning (left figure) and at the end of simulation (right figure). Colors of the nodes represent their phases. Green nodes $1,2,3$ and red nodes $4,5$ belong to two different clusters. For the right figure, blue connections denote intra-cluster links {\color{black}whose} coupling strengths converge to a constant value. Light-gray links correspond to the oscillating inter-cluster couplings {\color{black}whose} quasiperiodic trajectories approach the invariant manifold defined by \eqref{endtor} (see also Fig.~\ref{Fig:5-coupling}).}
\label{Fig:5-ba}
\end{figure}

\begin{figure}[!ht]
\center
\begin{overpic}[width=0.5\textwidth]{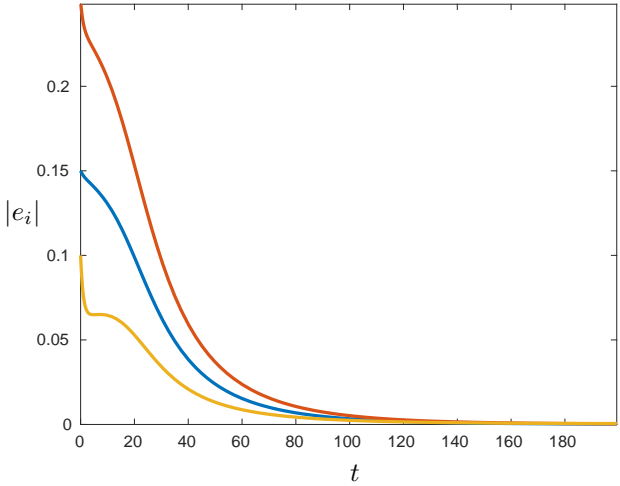}
 \put (52,0) {$\displaystyle t$}
 \put (2,38) {$\displaystyle |e_i|$}
\end{overpic}
\caption{(Example~1) Evolution of absolute values of the phase errors $e_i$, $i=\overline{1,N}$ within clusters. Convergence of the errors to zero corresponds to the asymptotic stability of the corresponding invariant manifold and the emergence of the two-cluster partition of the network.}
\label{Fig:5-errors}
\end{figure}

\begin{figure}[!ht]
\center
\begin{overpic}[width=0.5\textwidth]{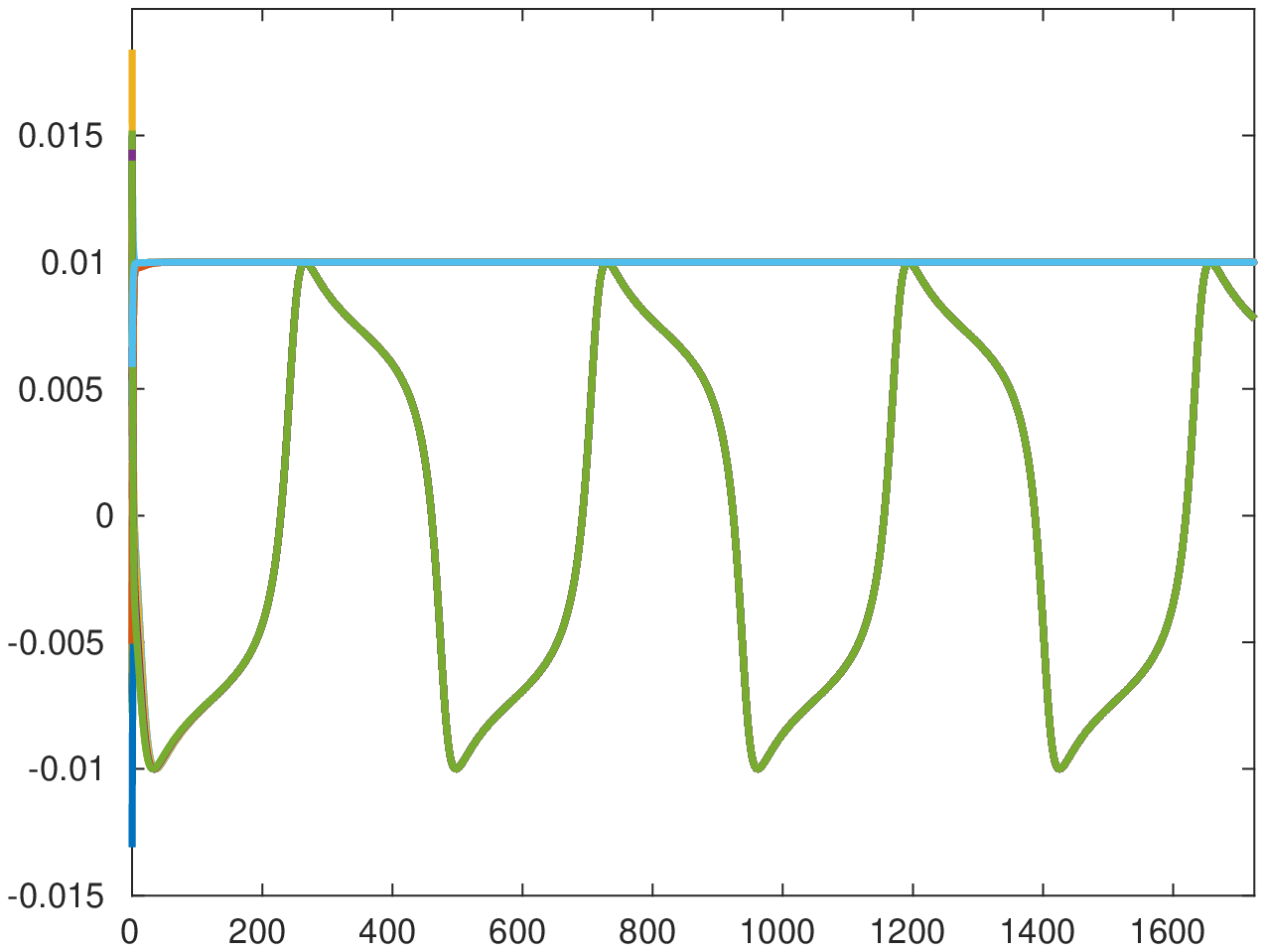}
 \put (52,0) {$\displaystyle t$}
 \put (2,33) {$\displaystyle k_{ij}$}
\end{overpic}
\caption{(Example~1) Evolution of coupling strengths $k_{ij}$.
The intra-cluster coupling strengths converge to the constant value $\frac{\tilde\mu}{\gamma}=0.01$ and the inter-cluster couplings converge to the invariant toroidal manifold \eqref{endtor} and exhibit quasiperiodic oscillations. Due to the choice of rationally independent natural frequencies, these oscillations are not periodic in time.}
\label{Fig:5-coupling}
\end{figure}

Consider the Kuramoto network $\Sigma(\tilde \mu, \mu)$ with $N=5$ all-to-all connected nodes, natural frequencies $w = (\frac{1}{2},\frac{1}{2},\frac{1}{2}, \frac{\sqrt{2}}{3}, \frac{\sqrt{2}}{3})^\top$, plasticity parameters $\tilde \mu = \mu=0.01$, $\gamma_1=1$, Hebbian learning rule $\Gamma(s)=\cos(s)$, and the desired two-cluster partition $\mathcal P = \mathcal P_1 \cup \mathcal P_2 = \{1,2,3\}\cup \{4,5\}$. The natural frequencies satisfy the condition (A1). Condition (A2) is satisfied thanks to the all-to-all connections between nodes. Directly calculating $c_{12}=2$, $c_{21}=3$, $c_{max}=3$, $c_{out}=12$, $\delta=1$, the conditions (A3):
\begin{equation*}
w_{min}-\mu\gamma^{-1}\delta c_{max}= \frac{\sqrt{2}}{3}-\frac{0.01}{1}\cdot 3 \approx 0.4614 > 0
\end{equation*}
and
\begin{equation*}
\begin{split}
4\frac{\mu}{\gamma^2} \delta \sqrt{c_{out}} \sum\limits_{\substack{s,r=\overline{1,m} \\ s\not = r}}{c_{sr}} \frac{w_{max}+\mu\gamma^{-1}\delta c_{max}}{w_{min}-\mu\gamma^{-1}\delta c_{max}} \approx 0.796 <1
\end{split}
\end{equation*}
are fulfilled. Following Remark 3, condition (A4) is satisfied due to the all-to-all connections and $\Gamma(0)=\cos(0)=1>0$. All conditions of Theorem 2 are satisfied. Simulation results for initial phases $\theta(0) = (\frac{\pi}{2}, \frac{\pi}{2}\mathord{+}0.15, \frac{\pi}{2}\mathord{+}0.25, 0, -0.1)^\top$ and random initial couplings $k_{ij}(0)\in [-0.015, 0.015]$, $i,j=\overline{1,5}$, $i\not = j$ are presented in Fig{\color{black}s}.~\ref{Fig:5-ba}, \ref{Fig:5-errors}, and \ref{Fig:5-coupling}.

\subsubsection{Example 2.}
Consider a network of $N=7$ Kuramoto oscillators \eqref{main1} with adjacency matrix
\begin{equation*}
A=
\left(
\begin{array}{ccc|cccc}
\ngrey 0 & \ngrey 1 & \ngrey 0 & 0 & 1 & 0 & 0 \\
\ngrey 0 & \ngrey 0 & \ngrey 1 & 0 & 0 & 0 & 1 \\
\ngrey 1 & \ngrey 0 & \ngrey 0 & 1 & 0 & 0 & 0 \\ \hline 
0 &  1 & 0 & \ngrey 0 & \ngrey 1 & \ngrey 0 & \ngrey 0 \\
0 &  1 & 0 & \ngrey 0 & \ngrey 0 & \ngrey 1 & \ngrey 0 \\
0 & 0 &  1 & \ngrey 0 & \ngrey 0 & \ngrey 0 & \ngrey 1 \\
 0 & 0 &  1 & \ngrey 1 & \ngrey 0 & \ngrey 0 & \ngrey 0 \\
\end{array}
\right)
\end{equation*}
and natural frequencies $w = \left(\frac{1}{2}, \frac{1}{2}, \frac{1}{2}, \frac{2}{\sqrt{5}}, \frac{2}{\sqrt{5}}, \frac{2}{\sqrt{5}}, \frac{2}{\sqrt{5}}\right)^\top$, plasticity parameters $\gamma = 0.2$, $\tilde \mu = \mu = 0.001$, Hebbian learning rule $\Gamma(s)=\cos(s)$, and the desired two-cluster partition $\mathcal P = \{1,2,3\} \cup \{4,5,6,7\}$. Condition (A1) of Theorem~\ref{thm_main} is satisfied thanks to the choice of $w$. 

Every node in cluster $\mathcal P_1$ has exactly one incoming link from the nodes of cluster $\mathcal P_2$, and vice versa. The resulting interconnection topology satisfies the condition (A2) from Theorem~\ref{thm_main} and the network satisfies (A3) with characteristics $c_{out}=7$, $c_{max}=c_{12}=c_{21}=1$, $\delta = 1$. Indeed,
\begin{equation*}
w_{min}-\mu\gamma^{-1}\delta c_{max}= \frac{1}{2}-\frac{0.001}{0.2}=0.495>0
\end{equation*}
and
\begin{equation*}
\begin{split}
4\frac{\mu}{\gamma^2} \delta \sqrt{c_{out}} \sum\limits_{\substack{s,r=\overline{1,m} \\ s\not = r}}{c_{sr}} \frac{w_{max}+\mu\gamma^{-1}\delta c_{max}}{w_{min}-\mu\gamma^{-1}\delta c_{max}} \approx 0.9615 <1.
\end{split}
\end{equation*}
All conditions of Theorem~\ref{thm_main} are satisfied. Asymptotic stability of the two-cluster formation can be concluded from the condition (A4) of Theorem~\ref{thm-stab}. For this purpose, the last nodes within each cluster, i.e., nodes $3$ and $7$, are chosen as $k_1$ and $k_2$, respectively. Since $\operatorname{sign}\Gamma(0)=\operatorname{sign}\cos(0)=1>0$, (A4) for the cluster $\mathcal P_1$ reads 
\begin{equation*}
A_1=\begin{pmatrix}
0&     1 &\vline& 0\\
0&     0 &\vline& 1\\ \hline
1 & 0 &\vline& 0 \\
\end{pmatrix}, \quad
\tilde A_1=\begin{pmatrix}
-1&     1\\
-1&     0
\end{pmatrix}, \quad
D_1^-=\begin{pmatrix}
1&     0\\
0&     1
\end{pmatrix}
\end{equation*}
so that
\begin{equation*}
\operatorname{Re} \lambda(\tilde A_1-D_1^-) = \operatorname{Re} \lambda \begin{pmatrix}
-2&     1\\
-1&     -1
\end{pmatrix}
=\operatorname{Re}\left(-\frac{3}{2}\pm\frac{\sqrt{3}}{2}i\right)<0.
\end{equation*}

\begin{figure}[!ht]
\begin{center}
\begin{overpic}[width=0.25\textwidth]{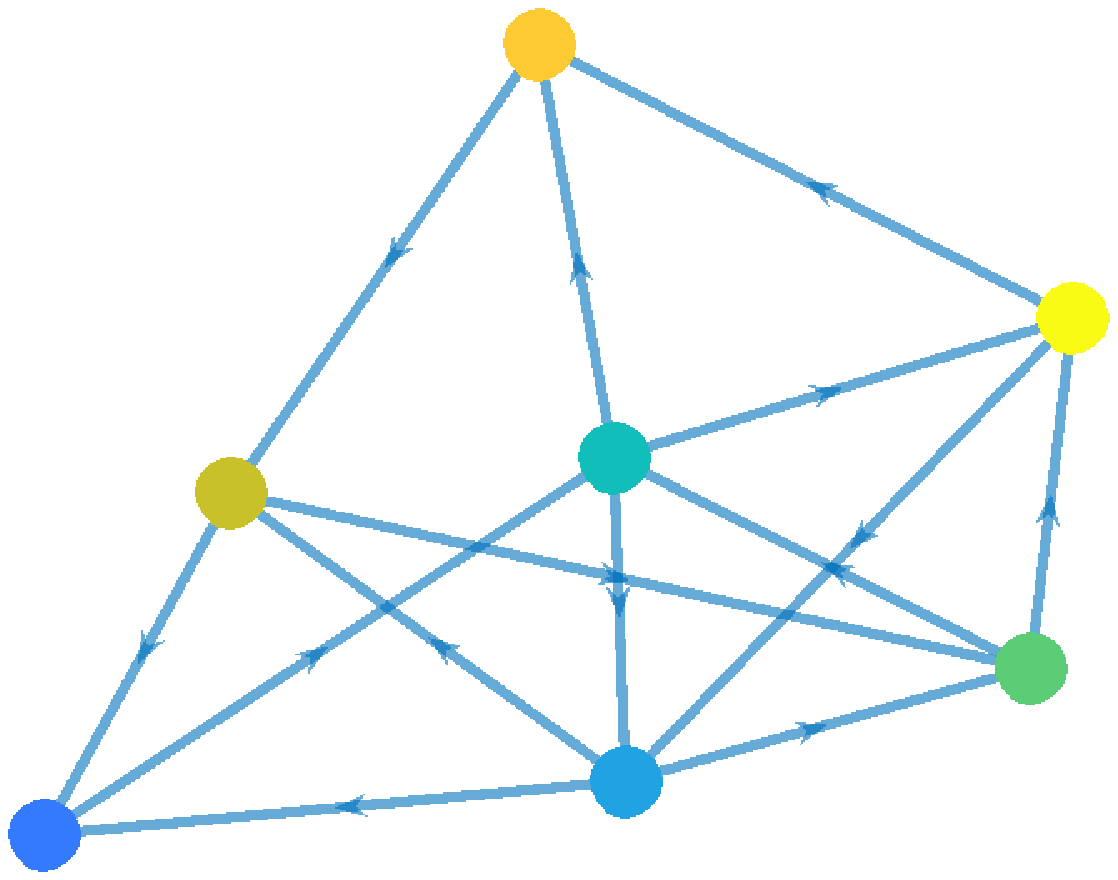}
 \put (15.3,10.5) {\tiny$\displaystyle 1$}
 \put (55.1,14.4) {\tiny$\displaystyle 2$}
 \put (54.4,36.6) {\tiny$\displaystyle 3$}
 \put (82.9,22.2) {\tiny$\displaystyle 4$}
 \put (28.2,34.0) {\tiny$\displaystyle 5$}
 \put (49,64.5) {\tiny$\displaystyle 6$}
 \put (85.8,45.9) {\tiny$\displaystyle 7$}
\end{overpic}\hspace{-6mm}
\begin{overpic}[width=0.25\textwidth]{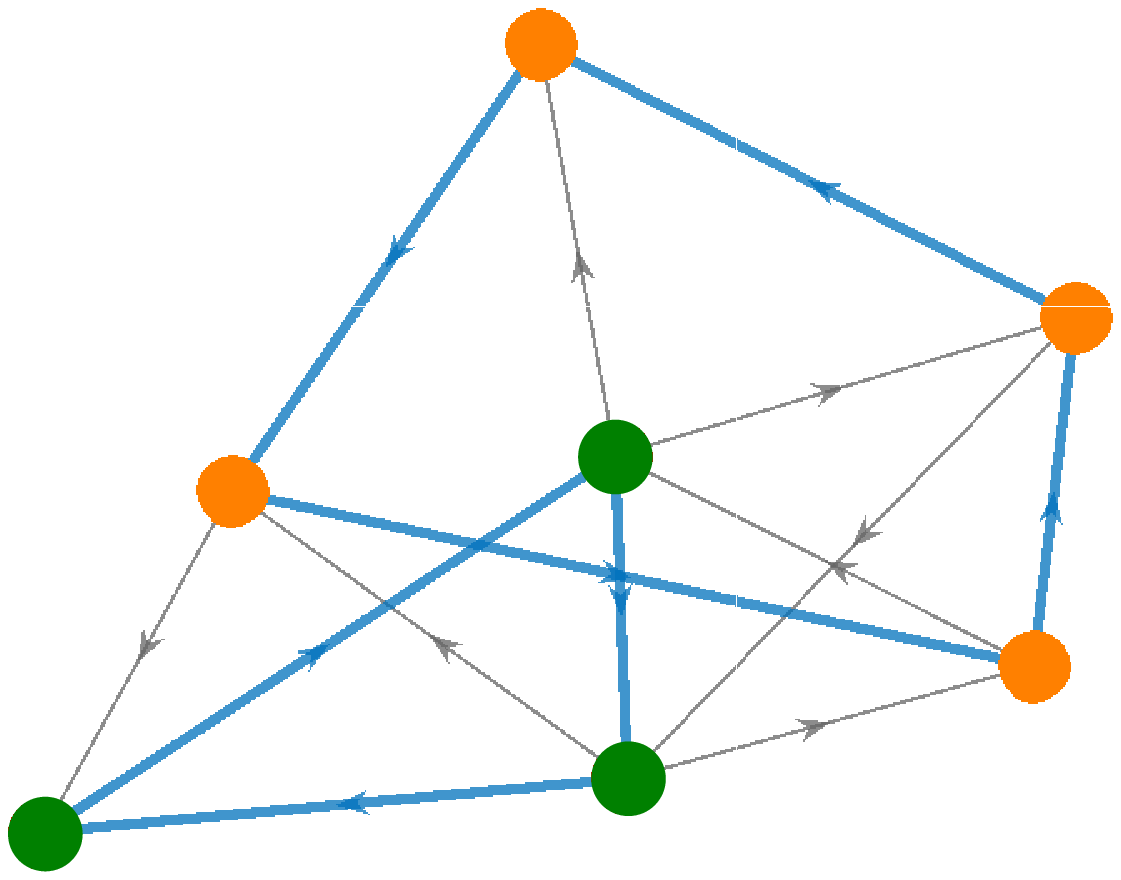}
 \put (15.3,10.5) {\tiny$\displaystyle 1$}
 \put (55.4,14.4) {\tiny$\displaystyle 2$}
 \put (54.6,36.4) {\tiny$\displaystyle 3$}
 \put (83.1,22.2) {\tiny$\displaystyle 4$}
 \put (28.2,34.0) {\tiny$\displaystyle 5$}
 \put (49.5,64.6) {\tiny$\displaystyle 6$}
 \put (85.99,45.9) {\tiny$\displaystyle 7$}
 \put (8,40) {\Large$\displaystyle \Rightarrow$}
\end{overpic}
\end{center}
\caption{(Example~2) Visualization of the graph $\mathcal G$ and coupling strengths at the beginning (left figure) and at the end of simulation (right figure). Colors of the nodes represent their phases. Green nodes $1,2$ and $3$ and orange nodes $4,5,6$, and $7$ belong to two different clusters. For the right figure, blue connections denote intra-cluster links {\color{black}whose} coupling strengths converge to a constant value. Light-grey links correspond to the oscillating inter-cluster couplings {\color{black}whose} quasiperiodic trajectories approach the invariant manifold defined by \eqref{endtor} (see also Fig.~\ref{Fig:3-coupling}).}
\label{Fig:topology}
\end{figure}

\begin{figure}[!ht]
\center
\begin{overpic}[width=0.5\textwidth]{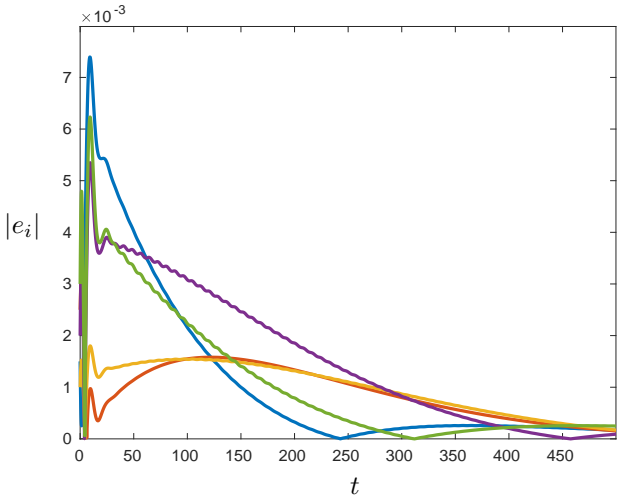}
 \put (52,0) {$\displaystyle t$}
 \put (2,38) {$\displaystyle |e_i|$}
\end{overpic}
\caption{(Example~2) Evolution of absolute values of the phase-errors $e_i$, $i=\overline{1,N}$ within clusters. All $e_i(t)\xrightarrow[]{t \to \infty} 0$, $i=\overline{1,N}$ \pfcom{that} corresponds to the asymptotic stability of the invariant toroidal manifold \eqref{endtor} and the emergence of two-cluster formation given by partition $\mathcal P$.}
\label{Fig:3-errors}
\end{figure}

\begin{figure}[!ht]
\center
\begin{overpic}[width=0.5\textwidth]{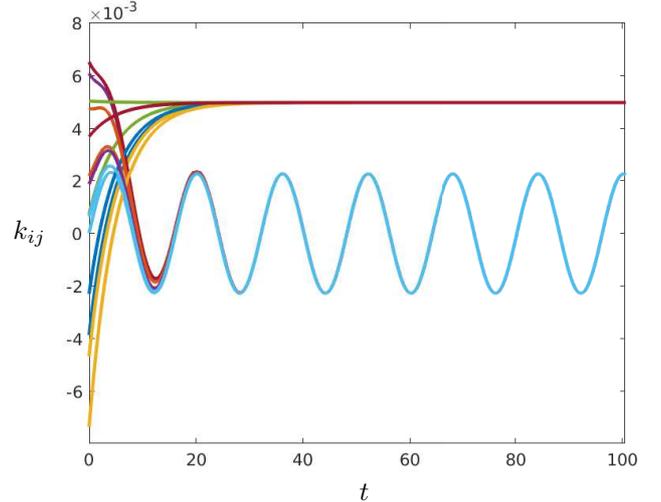}
 \put (52,0) {$\displaystyle t$}
 \put (2,38) {$\displaystyle k_{ij}$}
\end{overpic}
\caption{(Example~2) Evolution of coupling strengths $k_{ij}$. The intra-cluster coupling strengths converge to the constant value $\frac{\tilde\mu}{\gamma}=0.005$ and the inter-cluster couplings converge to the invariant toroidal manifold \eqref{endtor} and exhibit quasiperiodic oscillations. Due to the choice of rationally independent natural frequencies, these oscillations are not periodic in time.}
\label{Fig:3-coupling}
\end{figure}

For the cluster $\mathcal P_2$, condition (A4) reads
\begin{equation*}
A_2=\begin{pmatrix}
0&     1& 0&\vline & 0\\
0&     0& 1&\vline & 0\\
0& 	   0& 0&\vline & 1\\\hline
1& 	   0& 0&\vline & 0\\
\end{pmatrix}, \quad
\tilde A_2=\begin{pmatrix}
-1&     1& 0\\
-1&     0& 1\\
-1&     0& 0
\end{pmatrix}, \quad
D_2^-=\begin{pmatrix}
1&     0& 0\\
0&     1& 0\\
0&     0& 1
\end{pmatrix},
\end{equation*}
thus
\begin{equation*}
\operatorname{Re} \lambda(\tilde A_2-D_2^-) = \operatorname{Re} \lambda \begin{pmatrix}
-2&     1& 0\\
-1&     -1& 1\\
-1&     0& -1
\end{pmatrix}
=\left[
\begin{array}{l}
\operatorname{Re}(-1\pm i)\\ \operatorname{Re}(-2)\end{array}\right.<0.
\end{equation*}
All conditions (A1)-(A4) of Theorem 2 are satisfied. Simulation results for initial phases $$\theta(0) = \left(\frac{\pi}{2}, \frac{\pi}{2}+ \frac{3}{20}, \frac{\pi}{2}+ \frac{1}{4}, \frac{\pi}{3}- \frac{1}{10}, \frac{\pi}{3}- \frac{2}{10}, \frac{\pi}{3}- \frac{3}{10}\right)^\top$$ and random initial couplings $k_{ij}(0)\in [-0.01, 0.01]$, $i,j=\overline{1,N}$, $i\not = j$ are presented in Figs.~\ref{Fig:topology}, \ref{Fig:3-errors}, and \ref{Fig:3-coupling}.

\pfcom{Numerical simulations} for both examples demonstrate the conclusions of Theorem 2 and show that starting from a vicinity of invariant toroidal manifolds, which correspond to the desired multi-cluster behavior of oscillators, the coupling strengths and phase-errors of the considered Kuramoto networks converge to these manifolds.

\section{Conclusion}

In this paper, sufficient conditions have been derived for the asymptotic stability of the invariant toroidal manifolds corresponding to the multi-cluster behavior of Kuramoto networks with adaptive coupling. These conditions evince a decisive role of intra-cluster links for the stability of clusters. The paper proposes a qualitative stability result which additionally requires a sufficiently small plasticity parameter $\mu$. Quantitative estimates of the admissible plasticity parameters and of the region of attraction of the invariant manifold employing Lyapunov-based methods are of high interest for future research. 

\bibliography{ifacconf}            
\end{document}